\documentclass[a4paper,fleqn,usenatbib]{mnras}

\usepackage{graphicx} 
\usepackage{longtable} 
\usepackage{lscape}
\usepackage{rotating}
\usepackage[english]{babel}
\usepackage{grffile}
\usepackage{color}
\usepackage{grffile}
\usepackage[T1]{fontenc}
\usepackage{ae}
\usepackage{aecompl}
\usepackage{amsmath}
\usepackage{amssymb}

\newcommand{\snia}{SN~Ia}
\newcommand{\sneia}{SNe~Ia}
\newcommand{\msun}{$M_{\sun}$}

\title[White dwarfs and fast optical transients]
      {White dwarf dynamical interactions and fast optical transients}

\author[E. Garc\'\i a--Berro et al.]
       {Enrique Garc\'\i a--Berro$^{1,2}$,
        Carles Badenes$^{3}$,  
        Gabriela Aznar--Sigu\'{a}n$^{1,2}$ and\newauthor
        Pablo Lor\'{e}n--Aguilar$^{4}$
\\
       $^1$Departament de F\'{i}sica, 
           Universitat Polit\`{e}cnica de Catalunya,
           c/Esteve Terrades, 5, 
           08860 Castelldefels, 
           Spain\\
       $^2$Institut d'Estudis Espacials de Catalunya, 
           Ed. Nexus-201, 
           c/Gran Capit\`a 2-4, 
           08034 Barcelona, 
           Spain\\
       $^3$Department of Physics and Astronomy and 
           Pittsburgh Particle Physics, Astrophysics and Cosmology Center,
           University of Pittsburgh,\\
           Pittsburgh, 
           PA 15260,
           USA\\
       $^4$School of Physics, 
           University of Exeter, 
           Stocker Road, 
           Exeter EX4 4QL, 
           UK}

\date{Accepted 2017 March 29; in original form 2016 March 6}

\pubyear{2015}

\begin{document}
\label{firstpage}
\pagerange{\pageref{firstpage}--\pageref{lastpage}}
\maketitle

\begin{abstract}
Recent advances in time-domain astronomy have uncovered a new class of
optical transients with timescales shorter than typical supernovae and
a  wide  range  of  peak luminosities.   Several  subtypes  have  been
identified within this broad  class, including Ca-rich transients, .Ia
supernovae,  and fast/bright  transients. We  examine the  predictions
from  a  state-of-the-art  grid of  three-dimensional  simulations  of
dynamical  white  dwarf interactions  in  the  context of  these  fast
optical   transients.   We   find   that   for  collisions   involving
carbon-oxygen or  oxygen-neon white  dwarfs the peak  luminosities and
durations of the light curves in our models are in good agreement with
the properties of  fast/bright transients.  When one  of the colliding
white dwarfs is made of helium  the properties of the light curves are
similar to  those of  Ca-rich gap  transients.  The  model lightcurves
from our white dwarf collisions are too slow to reproduce those of .Ia
SNe, and too fast to match any normal or peculiar Type Ia supernova.
\end{abstract}

\begin{keywords}
Supernovae: general  -- Supernova remnants  -- Stars: white  dwarfs --
Hydrodynamics
\end{keywords}

%-----------------------------------------------------------------------

\section{Introduction}
\label{sec:Intro}

\begin{figure*}
\begin{center}
  \includegraphics[width=0.45\textwidth]{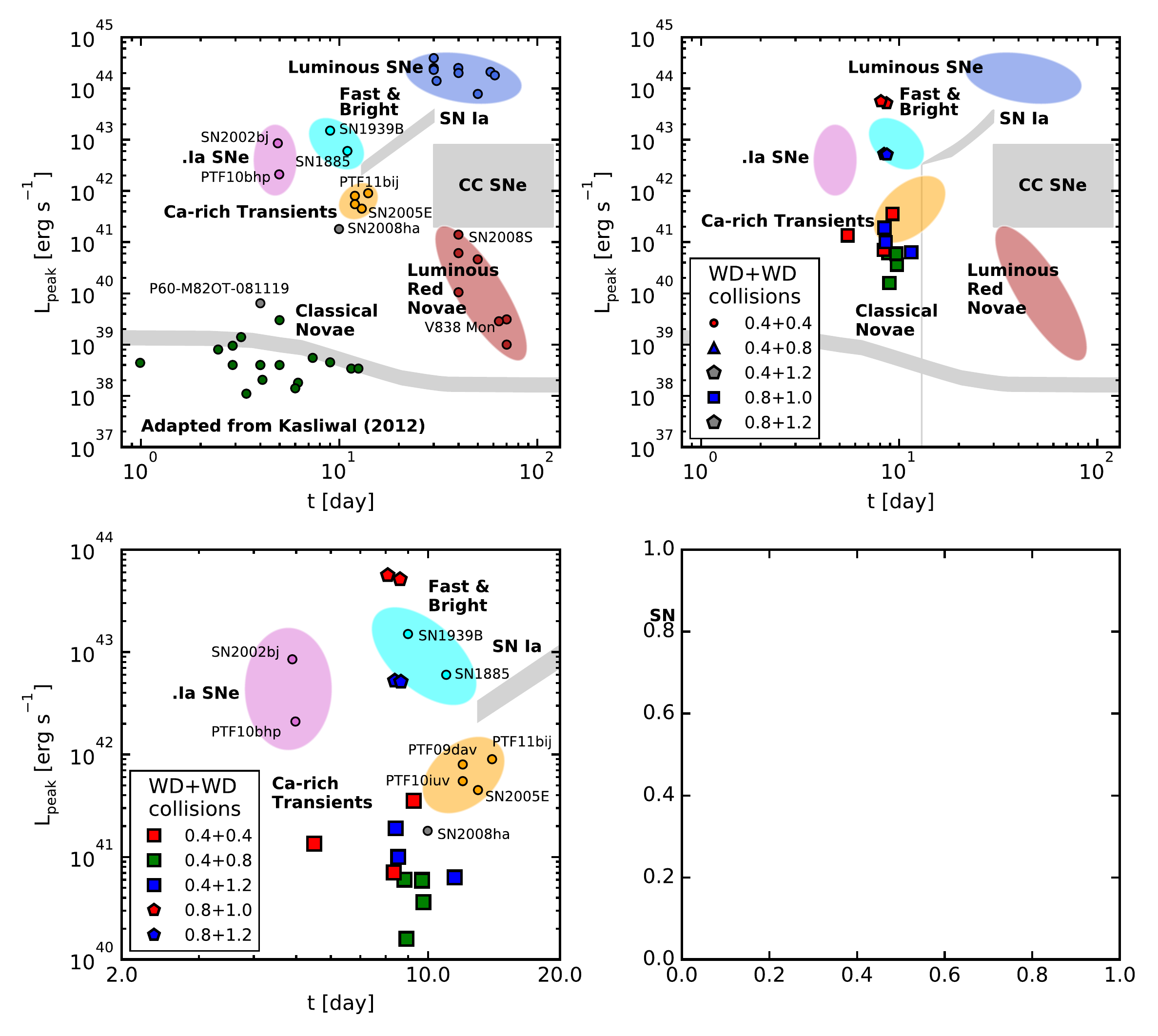}
  \includegraphics[width=0.46\textwidth]{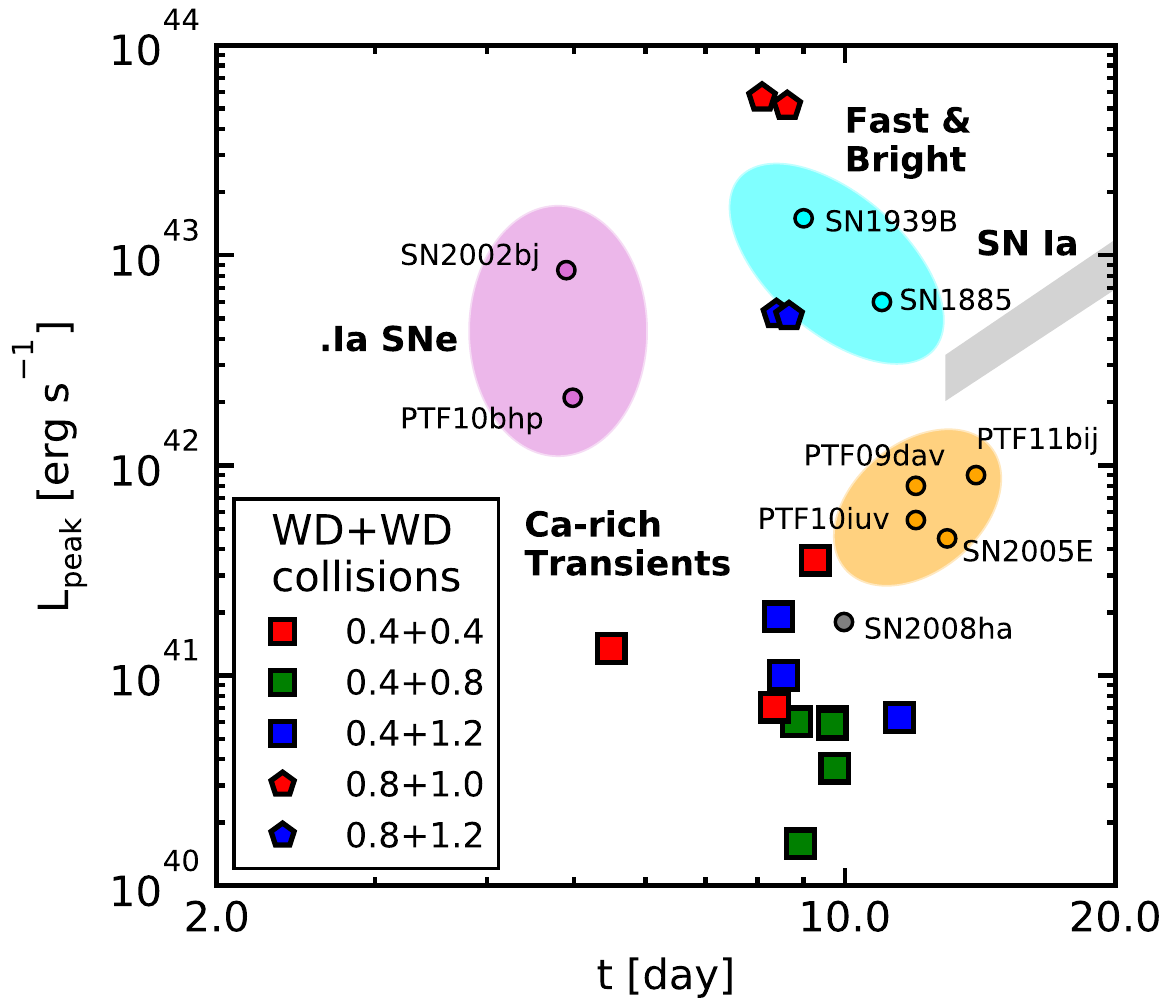}
\caption{Left  panel:   peak  luminosity   as   a   function  of   the
  characteristic timescale  for several optical transients,  with some
  relevant subclasses  highlighted in  color --- adapted from Kasliwal
  (2012).  Right  panel: location of several  simulations of colliding
  white dwarfs, for different  initial conditions, masses and chemical
  compositions  of  the interacting  white  dwarfs,  overlayed on  the
  observational subclasses discussed in the text.
\label{fig:gap}}
\end{center}
\end{figure*}

The advent  of photometric  surveys combining wide  field capabilities
with a fast cadence of visits, like PTF \citep{Law2009} and Pan-STARRS
\citep{Kaiser2002},  has led  to significant  advances in  time-domain
astrophysics. In particular, these surveys  have uncovered a new class
of astronomical  transients with  timescales faster than  the $\sim$20
days typical of Type Ia supernovae  (\sneia), and a wide range of peak
luminosities,   which   we   will    broadly   categorize   as   `fast
transients'. Because this is an emerging field, the full observational
range  of fast  transients has  yet to  be characterized,  but several
sub-classes have already been described  in the literature --- see the
left panel  of Figure~\ref{fig:gap}, adapted  from \cite{Kasliwal12a}.
The  so-called `fast  and bright'  transients \citep{Perets2011}  were
identified  partly  through  historical observations  of  SN~1885  and
SN~1939B.    They  are   as   luminous  as   normal  \sneia\   ($L\sim
10^{43}$~erg~s$^{-1}$, or  $M_{V} \sim  -18$), but have  timescales of
$\sim 10$~days, faster  than even the faintest,  more rapidly evolving
\sneia.      The     somewhat      awkwardly     named     .Ia     SNe
\citep{Bildsten2007,Poznanski,Kasliwal2010} have  luminosities similar
to fast/bright  transients, but even  faster timescales of only  a few
days.  Ca-rich  gap transients have timescales  similar to fast/bright
transients,  but are  an order  of magnitude  fainter at  peak ($L\sim
10^{42}$~erg~s$^{-1}$, or  $M_{V} \sim -16$), and  show strong nebular
features      from     Ca      in     their      late-time     spectra
\citep{Filippenko03,Perets10,Kasliwal12b}.   Finally,  there are  some
fast transients  like SN2008ha  that do  not seem to  fit into  any of
these categories \citep{Foley10}.  At present,  it is still unclear to
what  degree  these  classes  are physically  distinct  or  internally
homogeneous, and  it is  likely that the  taxonomy of  fast transients
will evolve in the near future.

Our  understanding of  the stellar  progenitors for  each subclass  of
these fast  transients is still  very limited.  The exception  are .Ia
SNe,  which were  theoretically  predicted  by \cite{Bildsten2007}  to
arise  from the  thermonuclear  detonation  of a  helium  shell on  an
accreting white  dwarf.  Their  observational counterparts  were later
identified by \cite{Kasliwal2010}, and  found to exhibit many features
in agreement with  this model, so this subclass has  at least a viable
progenitor scenario ---  but see \cite{Shen10} for  a discussion.  The
situation   is   less   clear    for   the   other   two   subclasses.
\cite{Perets2011}  proposed  a  white  dwarf  origin  for  fast/bright
transients, based on  the old stellar environments  around SN~1885 and
SN~1939B, the  low ejecta masses  inferred from the  short timescales,
and the  lack of detectable X-ray  emission after more than  130 years
from SN~1885.

A  number  of  competing  models  are  being  considered  for  Ca-rich
transients, including gravitational core  collapse of stripped massive
stars \citep{Kawabata10}, the coalescence of a binary system made of a
neutron  star  and a  white  dwarf  \citep{Metzger12,Sell15}, and  the
merger of  two white dwarfs, one  of them with a  carbon-oxygen (or an
oxygen-neon)   core    and   the    second   one   made    of   helium
\citep{Perets10,Kasliwal12b}. The first of these scenarios, namely the
explosion of a  rather massive star ($M\ga 8\,M_{\sun}$),  seems to be
in contradiction with  the prevalence of early-type  host galaxies for
these  outbursts, and  with  the fact  that  Ca-rich transients  occur
predominantly  in  galaxies with  no  apparent  signs of  recent  star
formation  \citep{Perets10,Lyman13}.    Moreover,  Ca-rich  transients
occur     predominantly    in     the    outskirsts     of    galaxies
\citep{Kasliwal12b,Lyman14,Foley15}, and their  locations do not trace
the stellar  light of their  host galaxies,  at odds with  what occurs
with other luminous transients  and supernovae \citep{Anderson}.  This
led to the  suggestion that these events are due  to explosions in old
white dwarfs belonging  to faint globular clusters  or dwarf galaxies,
which are difficult to detect --- see, for instance, \cite{Waldman11}.
However, \cite{Lyman14} obtained deep VLT images of two members of the
class (SN~2005E and  2012hn) and found no evidence  for any underlying
stellar population.   More recently, \cite{Lyman16} analyzed  HST data
of  five  objects  of  the  class  (SN~2001co,  SN~2003dg,  SN~2003dr,
SN~2005E,  and SN~2007ke)  and corroborated  their previous  findings.
This  failure to  identify faint  and dense  stellar systems  --- like
globular clusters, or dwarf galaxies  --- as the underlying population
does not  necessarily mean that  these transients could not  have been
originated in  other old stellar  populations --- like  stellar haloes
--- which would be too faint to  be detected, even with image stacking
--- see  \cite{Perets2014}, and  references therein.   Nevertheless, a
cautionary  remark is  in  order  here, as  the  rate  at which  these
collisions occur in typical old  stellar populations is expected to be
small  \citep{Perets2014}.  In  summary, the  issue of  the origin  of
Ca-rich transients remains unresolved.

An  interesting  possibility  is   that  the  progenitors  of  Ca-rich
transients are not formed at large offsets from their parent galaxies,
but instead that they have  travelled large distances because of their
large peculiar velocities \citep{Lyman14}.  This, in turn, could point
towards a dynamical interaction of  a white dwarf with another compact
object, either a black hole  \citep{Sell15, MacLeod16}, a neutron star
\citep{Metzger12}, or a white dwarf \citep{Brown11, Foley15}, in which
the  system  acquires  a  sufficiently  high  kick.   Since  dynamical
interactions (either  mergers or collisions) between  two white dwarfs
should be  more common than  those between a  black hole or  a neutron
star and  a white  dwarf, the  former channel  should be  more likely.
Indeed,  we  know  from  the  distribution  of  radial  velocities  in
multi-epoch observations  of Galactic  white dwarfs  that coalescences
and/or    collisions    happen    at     a    rate    comparable    to
\sneia\  \citep{BadenesMaoz2012,Maoz2016},  which  imples  that  their
observational counterparts,  whatever they may be,  must be relatively
common. In  this scenario, the  explosion could  be the result  of the
merger of  double white dwarf  binary system where the  components are
brought  together by  gravitational  wave radiation,  or  it could  be
triggered  by a  stellar close  encounter, perhaps  prompted by  Kozai
oscillations           in           a          triple           system
\citep{Thompson11,Katz2012,Antognini14} ---  see Sect.~\ref{wddi}.  In
either  case,  the  expectation  is  that  for  some  interactions  an
explosion will occur at the base of the accreted buffer that is formed
after the disruption of the less massive star.

Here we study the possibility that some of these fast transients arise
from  dynamical  interactions  between   two  white  dwarfs  that  are
triggered either by random encounters in dense stellar environments or
by the coalescence of binary  pairs.  Although the physical mechanisms
driving white  dwarf mergers and  collisions are quite  different, the
overall  properties  of  the  dynamical interaction  are  similar  for
similar masses of the interacting  white dwarfs.  More importantly, in
both cases variable  amounts of nickel are synthesized  in the ensuing
thermonuclear flash, thus powering quite different light curves.  This
could  help  to explain  the  observed  features of  some  transients.
Perhaps the single major, but significant, difference of between white
dwarf mergers and collisions is that in white dwarf mergers the debris
region surrounding the central compact  object consists of a Keplerian
disk,  while  in white  dwarf  collisions  the  disk, if  present,  is
embedded in a  shroud of high-velocity material.  Finally,  it is also
worth  mentioning that  in some  white dwarf  collisions the  velocity
field of the  remnant of the interaction is  highly asymmetric.  Thus,
owing to linear momentum conservation, the remnants of the interaction
can  move   at  considerable  speeds.    This  may  explain   why  the
distribution of the locations of Ca-rich transients is strongly skewed
to  very large  galactocentric offsets  \cite{Anderson}, unlike  other
luminous transients and SNe.

In this paper,  we explore the predictions from a  model grid of white
dwarf dynamical  interactions \citep{Gabriela1} in the  context of the
newly identified subclasses  of fast transients.  We  have chosen this
model grid  because it is  the most consistent suite  of calculations,
although  the results  of other  sets  of calculations  of merging  or
colliding  white  dwarfs  are  quantitatively similar.   Our  work  is
organized  as follows.   In Sect.~\ref{wddi}  we review  the different
sets  of  calculations  of  white dwarf  dynamical  interactions.   In
Sect.~\ref{sec:Lpeak} we  discuss the light curves  resulting from our
dynamical white  dwarf interaction models.   We show that when  one of
the colliding  white dwarfs  is made  of carbon  and oxygen,  the peak
luminosities   and  timescales   agree  with   those  of   fast/bright
transients, and when  one of the white dwarfs has  a helium core, they
are similar to  those of Ca-rich transients.   In Sect.~\ref{sec:X} we
argue  that  the calcium  yields  in  our  models, and  their  spatial
distribution, are  also in reasonable agreement  with the observations
of  some Ca-rich  transients.   Finally,  in Sect.~\ref{sec:concl}  we
summarize our findings and we describe our conclusions.

\section{White dwarf dynamical interactions}
\label{wddi}

White  dwarf dynamical  interactions  can be  classified  in two  main
categories:  white dwarf  mergers, and  white dwarf  collisions. White
dwarf  mergers are  driven  by the  emission  of gravitational  waves,
whereas white dwarf collisions occur  when two white dwarfs pass close
to each  other, and gravitational  focusing brings them so  close that
the  less gravitationally  bound star  transfers mass  to the  massive
member of  the system. In  both cases the  result is that  the lighter
white dwarf  is destroyed during  the interaction and its  material is
accreted onto the heavier white dwarf.

In white dwarf mergers the accretion episode occurs when the secondary
star overflows  its Roche  lobe, while in  white dwarf  collisions the
secondary usually  begins transferring  mass at periastron.   Once the
dynamical interaction between the two white dwarfs begins, the process
of disruption of the lightest member of the system is accelerated, due
to  the inverse  dependence  of  the radius  of  white  dwarfs on  the
mass. Depending  on the masses of  the coalescing white dwarfs  and on
the  corresponding mass  ratio the  process can  be very  fast, and  a
common result  is that  the disruption  takes place  in a  few orbital
periods for the  case of white dwarf mergers, and  on short timescales
for white dwarf collisions in which the low-mass member of the pair is
captured in  a highly eccentric  orbit, typically  in a few  orbits as
well.  In both cases it turns out that the material transferred to the
heavier star is compressed and heated  to such an extent that, in some
cases,  the conditions  for a  detonation  are met.  This, of  course,
depends on the initial conditions of the dynamical interaction, and on
the masses and chemical compositions of the intervening white dwarfs.

White dwarf mergers have been  extensively studied during the last few
years, because  they may  possibly account for  a sizable  fraction of
normal Type Ia supernova  explosions. The pioneering Smoothed Particle
Hydrodynamics  (SPH) calculations  of \cite{Benz89a,  Benz89b, Benz90}
paved the  way to what  now is a  mature research field.   These early
calculations were  followed some time  later by more  accurate studies
\citep{RS95, Segretain1997}.  However, progress was slow in subsequent
years, until new sets of numerical calculations overcoming some of the
limitations   of    the   previous   simulations    became   available
\citep{Guerrero, Yoon}.   In these  new calculations  enhanced spatial
resolutions,  improved  prescriptions  for the  artificial  viscosity,
updated nuclear reaction rates, and  realistic equations of state were
employed to cover a wider range of masses and chemical compositions of
the  merging  white  dwarfs.   The   most  recent  sets  of  numerical
calculations  of merging  white  dwarfs  \citep{PabloM, Pak10,  Pak11,
Rosswog11, Pak12a, Pak12b, Rosswog12, Raskin12, Pak13a, Pak13b, Zhu13,
Rosswog14, Raskin14a,  Raskin14b, Zhu15, Pak15,  Nomoto15a, Nomoto15b,
Rosswog15} have  come to an  agreement about the evolution  during the
merger  episode and  the characteristics  of the  merged remnant.   In
particular, there is  a general consensus that little  mass is ejected
from  the system  during the  interaction, and  that a  violent merger
(that is, a  merger in which a prompt explosion  takes place) does not
occur except in those cases in  which two rather massive ($\sim 0.9 \,
M_{\sun}$)    carbon-oxygen   white    dwarfs   of    similar   masses
merge. Moreover,  in those mergers  that do not experience  a powerful
detonation, and consequently are not  totally disrupted, about half of
the mass  of the secondary  is accreted onto  the primary star  of the
pair,  while the  rest of  the mass  forms a  debris region  which for
unequal  mass mergers  is  a Keplerian  rotating disk.   Additionally,
variable  amounts   of  nickel  and  intermediate-mass   elements  are
synthesized   and,  consequently,   the  peak   luminosities  of   the
corresponding  light   curves  span  a  considerable   range  ---  see
Fig.~\ref{fig:gap}. The resulting nucleosynthesis  is sensitive to the
mass ratio  and to  the chemical composition  of the  coalescing white
dwarfs.

\begin{figure}
\begin{center}
  \includegraphics[width=0.95\columnwidth,clip=true]{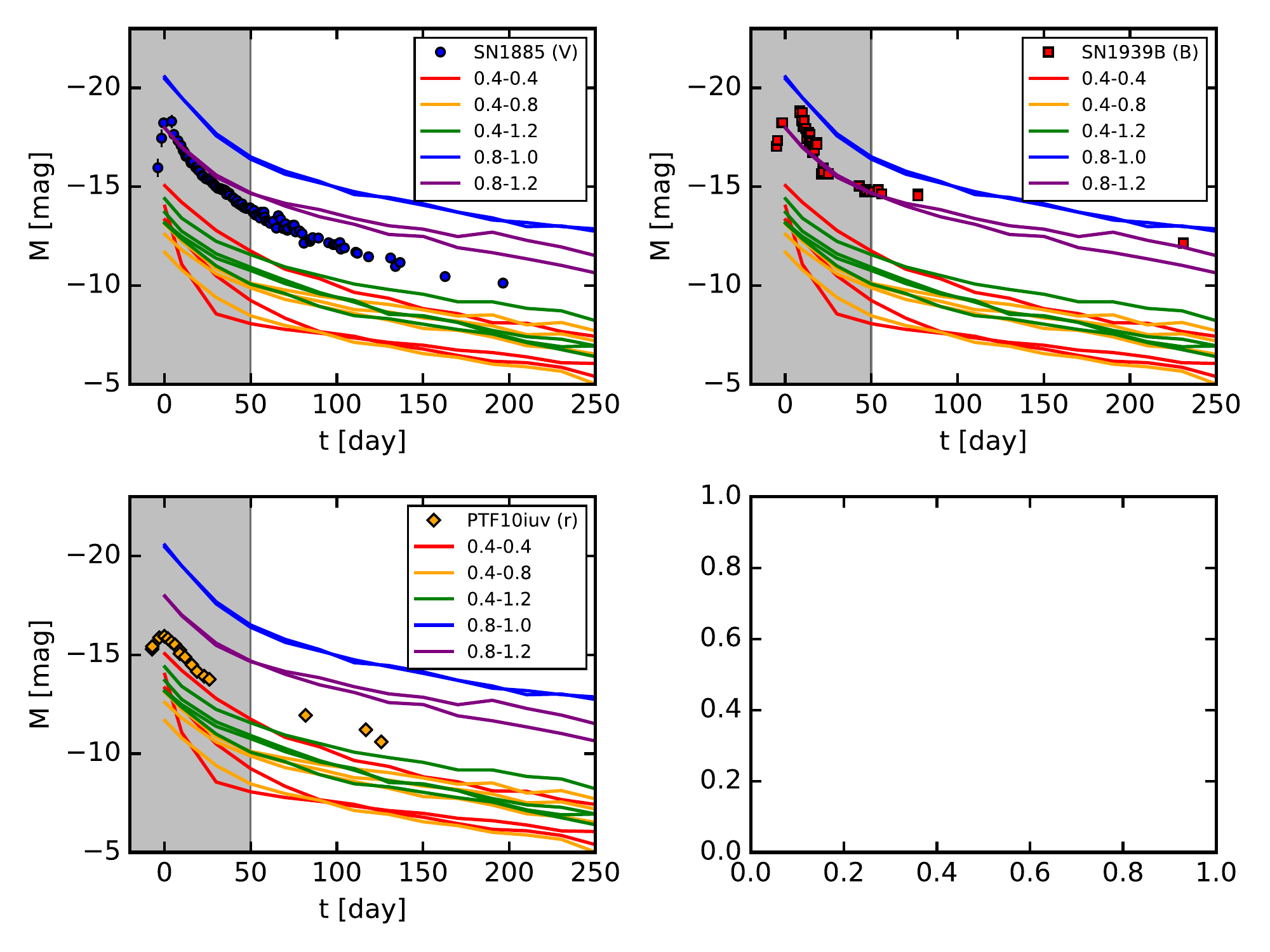}\\
  \includegraphics[width=0.95\columnwidth,clip=true]{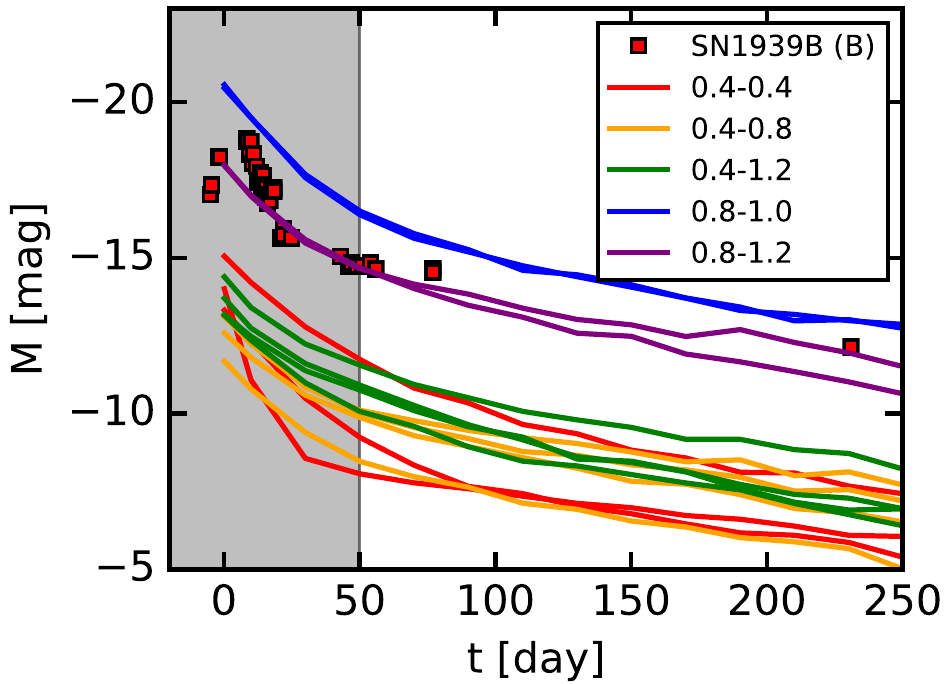}\\
  \includegraphics[width=0.95\columnwidth,clip=true]{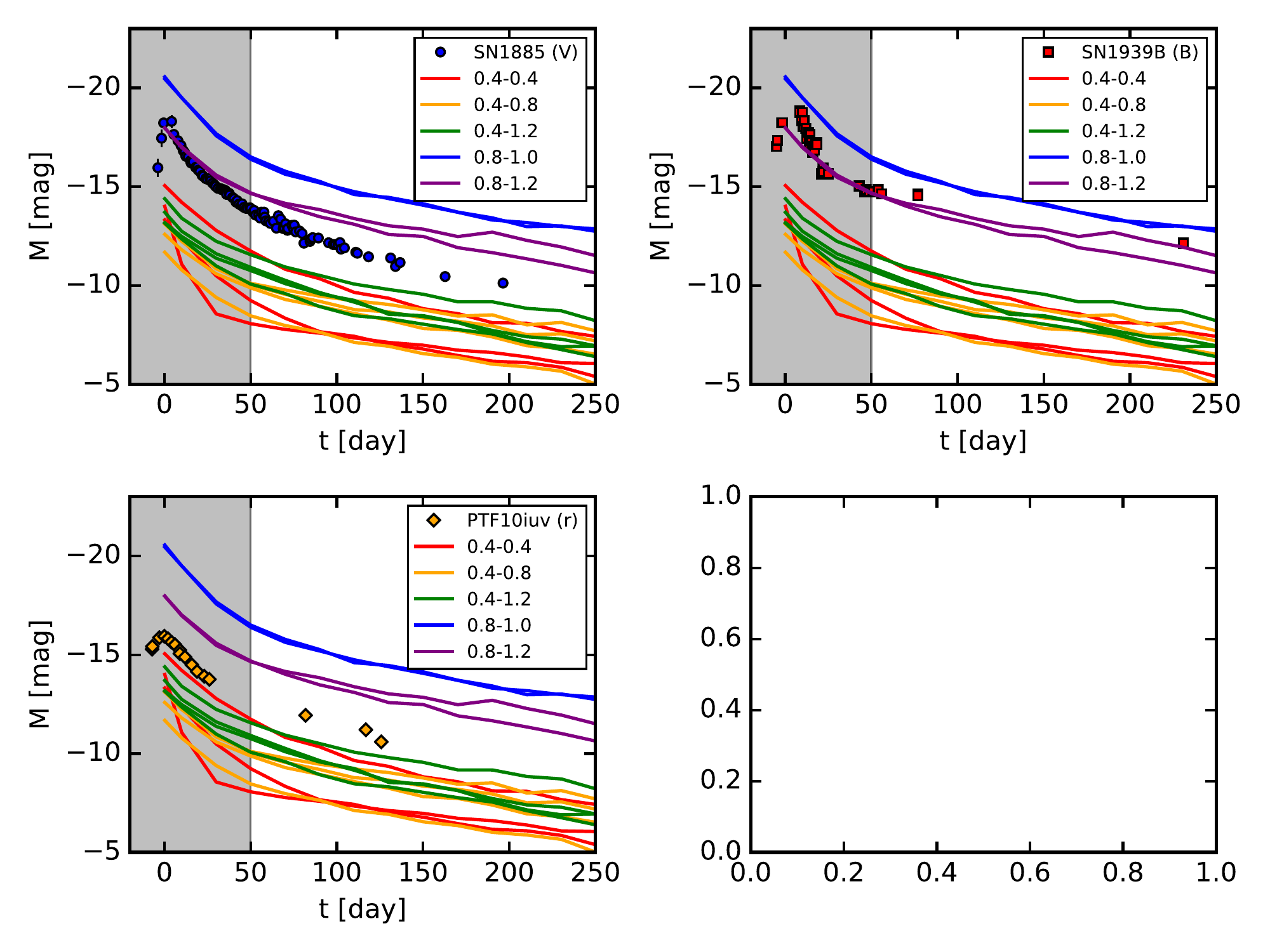}\\
  \caption{Light curves  of SN~1885, SN~1939B, and  PTF~10iuv --- from
    \citet{Perets2011}  and \citet{Kasliwal12b}  ---  compared to  the
    predicted light curves  from our models. The regime  below 50 days
    after   peak,  where   our  calculated   light  curves   are  more
    approximative, is shaded in gray.   The lightcurve of PTF10iuv has
    been  corrected  for host  galaxy  contamination  as described  in
    \citet{Kasliwal12b}.
\label{fig:LC}}
\end{center}
\end{figure}

The study of white dwarf collisions was also pioneered by \cite{Benz},
who employed a  SPH code to simulate the collisions  of two systems of
white  dwarfs of  masses $0.6\,  M_{\sun}$ and  $0.6\, M_{\sun}$,  and
$0.9\, M_{\sun}$ and $0.7\,  M_{\sun}$, respectively.  For two decades
this study  remained the only  one to examine white  dwarf collisions.
The  first recent  study of  white dwarf  collisions was  performed by
\cite{Rosswog2009}.  However,  \cite{Rosswog2009} limited  their study
to head-on collisions, although  they investigated the characteristics
of  the dynamical  interaction  for several  masses  of the  colliding
carbon-oxygen white dwarfs.   Almost simultaneously, \cite{Raskin2009}
examined the collision of two  otherwise typical white dwarfs of equal
masses  ($0.6\, M_{\sun}$)  for different  impact parameters  and mass
resolutions.  In  the same vein, \cite{Pablo}  explored the collisions
of  one mass  pair (of  unequal  masses, $0.6\,  M_{\sun}$ and  $0.8\,
M_{\sun}$) for various impact  parameters. Later on, \cite{Raskin2010}
expanded their previous  study to encompass pairs  of different masses
and collisions with different impact parameters.  In all these sets of
simulations  state-of-the-art  SPH  codes  were  employed.   Moreover,
reliable  equations  of state,  and  large  numbers of  SPH  particles
(ranging  from  $4\times  10^5$   to  $2\times  10^6$)  were  adopted.
Finally,  \cite{Domingo} studied  the  head-on collision  of two  twin
white dwarfs of  mass $0.7\, M_{\sun}$, but although they  used a very
high mass resolution, the simulations were not three-dimensional.  The
only  studies in  which  a SPH  code  was not  employed  are those  of
\cite{Hawley2012},  {\cite{Kushnir2013}}  and \cite{Papish}.   In  all
cases   the   Eulerian   adaptive    grid   code   FLASH   was   used.
\cite{Hawley2012} studied two twin  pairs of masses $0.64\, M_{\sun}$,
and $0.81\, M_{\sun}$, whereas \cite{Papish} considered several masses
of  the  colliding  white  dwarfs, from  $0.6\,  M_{\sun}$  to  $0.8\,
M_{\sun}$,    and   \citet{Kushnir2013}    performed   two-dimensional
hydrodynamical  simulatiosn  of  zero-impact-parameter  collisions  of
white dwarfs  with masses  between $0.5$ and  $1.0\, M_{\sun}$.  It is
noteworthy that  in all these  simulations the colliding  white dwarfs
were made of carbon and oxygen.

\begin{table*}
\caption{Calcium  masses  (in  solar   masses)  for  the  white  dwarf
  collisions  of  Aznar--Sigu\'an et  al.  (2013)  that result  in  an
  explosion.}
\label{tab:calcium}
\centering
\begin{tabular}{cccccc}
\hline
$M_1+M_2$    & $v_{\rm ini}$ & $\Delta y$ & $M_{\rm Ca}$ & $M_{\rm Ni}$ & $M_{\rm Fe}$ \\
$(M_{\sun})$ & (km~s$^{-1}$) & $(R_{\sun})$ & $(M_{\sun})$ & $(M_{\sun})$ & $(M_{\sun})$\\
\hline
0.4+1.2 & 150 & 0.3 & $7.3\times 10^{-3}$ & $8.4\times 10^{-4}$ & $1.6\times 10^{-2}$ \\
0.4+1.2 & 100 & 0.3 & $7.0\times 10^{-3}$ & $2.2\times 10^{-3}$ & $2.0\times 10^{-2}$ \\
0.4+1.2 & 100 & 0.4 & $7.3\times 10^{-3}$ & $9.4\times 10^{-4}$ & $1.3\times 10^{-2}$ \\
0.4+1.2 &  75 & 0.4 & $7.4\times 10^{-3}$ & $1.2\times 10^{-3}$ & $1.5\times 10^{-2}$ \\
0.4+0.8 & 150 & 0.3 & $8.5\times 10^{-3}$ & $2.0\times 10^{-4}$ & $5.4\times 10^{-3}$ \\
0.4+0.8 & 100 & 0.3 & $8.4\times 10^{-3}$ & $7.6\times 10^{-4}$ & $1.7\times 10^{-2}$ \\
0.4+0.8 & 100 & 0.4 & $8.5\times 10^{-3}$ & $5.0\times 10^{-4}$ & $9.0\times 10^{-3}$ \\
0.4+0.8 &  75 & 0.4 & $8.2\times 10^{-3}$ & $8.0\times 10^{-4}$ & $1.2\times 10^{-2}$ \\
0.4+0.4 & 100 & 0.3 & $1.9\times 10^{-2}$ & $4.7\times 10^{-3}$ & $2.9\times 10^{-2}$ \\
0.4+0.4 &  75 & 0.3 & $1.8\times 10^{-3}$ & $8.8\times 10^{-4}$ & $1.7\times 10^{-2}$ \\
0.4+0.4 &  75 & 0.4 & $2.0\times 10^{-2}$ & $1.6\times 10^{-3}$ & $1.9\times 10^{-2}$ \\
0.4+0.2 &  75 & 0.3 & $1.6\times 10^{-3}$ & $1.3\times 10^{-9}$ & $3.5\times 10^{-7}$ \\
\hline
0.8+1.2 & 100 & 0.3 & $1.9\times 10^{-2}$ & $6.3\times 10^{-2}$ & $3.6\times 10^{-3}$ \\
0.8+1.2 &  75 & 0.4 & $1.9\times 10^{-2}$ & $6.3\times 10^{-2}$ & $3.6\times 10^{-3}$ \\
0.8+1.0 & 100 & 0.3 & $6.3\times 10^{-2}$ & $7.2\times 10^{-1}$ & $1.6\times 10^{-2}$ \\
0.8+1.0 &  75 & 0.4 & $6.5\times 10^{-2}$ & $7.3\times 10^{-1}$ & $1.6\times 10^{-2}$ \\
\hline
\end{tabular}
\end{table*}

Inspired in part by the observational discoveries previously mentioned
in  Sect.~\ref{sec:Intro},  and  by  the  recent  theoretical  efforts
discussed in the previous paragraphs, \cite{Gabriela1} computed a grid
of white  dwarf collisions,  considering a  large interval  of initial
conditions and a broad range  of masses and core chemical compositions
of the interacting white dwarfs.   This suite of simulations, together
with  that of  \cite{Raskin2010} are  the most  comprehensive sets  of
calculations  of   white  dwarf  collisions  performed   so  far,  and
supplement previous calculations of  this kind. However, the advantage
of the  simulations of \cite{Gabriela1}  over the rest  of theoretical
calculations is that they explored the role of the core composition of
the  colliding  white dwarfs.   In  particular,  they computed  in  an
homogenous way a suite of collisions in which one of the components of
the  pair  of white  dwarfs  was  made  of helium,  carbon-oxygen,  or
oxygen-neon,  whereas the  second member  of  the system  was made  of
helium or  carbon-oxygen.  This particular feature  of the simulations
of \cite{Gabriela1} is important, because it opens new and interesting
possibilities  for   observational  counterparts,  that  need   to  be
explored.  As a  matter of fact, the  calculations of \cite{Gabriela1}
reveal that, independently of the core compositions of the interacting
white  dwarfs, the  result of  the dynamical  interaction has  several
remarkable characteristics.  The first of  them is that the remnant of
the  interaction is  surrounded  by  a shroud  of  material moving  at
considerable  speeds.    The  second  is  that   variable  amounts  of
intermediate-mass elements are synthesized,  depending on the chemical
composition of the colliding white  dwarfs.  The amount of radioactive
nickel  resulting in  these  simulations spans  a considerable  range,
resulting  in very  different peak  luminosities of  the corresponding
light curves.  Finally, because of the large asymmetry of the velocity
field resulting from the dynamical  interaction, the debris region can
move at sizable speeds, imparting an asymmetric kick to the ejecta. In
summary,  all these  characteristics of  the theoretical  calculations
deserve further attention,  and consequently it is  worth studying the
possibility that they  could explain, at least partially,  some of the
observed properties of fast optical transients.

\section{Peak luminosities, decay times, and light curves}
\label{sec:Lpeak}

\cite{Gabriela1} showed that in some  cases the outcome of white dwarf
close  encounters  could be  a  direct  collision  in which  only  one
catastrophic episode  of mass  transfer between the  less-massive star
and   the  massive   white   dwarf  occurs,   or  lateral   collisions
characterized   by  several   more   gentle  mass-transfer   episodes.
Moreover, they also  demonstrated that in a significant  number of the
cases, the material of the disrupted  star is compressed and heated to
such an  extent that the  conditions for detonation are  met, reaching
peak temperatures of the order of $10^9$~K, and even larger for direct
collisons, for  which temperatures larger than  $10^{10}$~K are easily
reached.  Accordingly, the masses  of $^{56}$Ni synthesized during the
detonation  span several  orders  of magnitude,  from $\sim  1.3\times
10^{-9}\, M_{\sun}$ to about $0.7\, M_{\sun}$.

\cite{Gabriela2}  computed the  light  curves for  the simulations  in
\cite{Gabriela1}.  The peak luminosities  of these interactions span a
wide  range,   between  $1.5\times  10^{43}$~erg~s$^{-1}$,   which  is
brighter than even the brightest \snia, and $2.7\times 10^{34}$, which
is fainter  than even  typical novae.  Obviously,  this wide  range of
peak luminosities is the result of the very different $^{56}$Ni masses
produced during the most violent  phases of the interactions. Although
\cite{Gabriela2} employed  an approximate method to  compute the light
curves,  their  corresponding  durations  can  be  determined  with  a
reasonable degree of accuracy. Specifically, \cite{Gabriela2} employed
the approximate  method of \cite{Kushnir13}, which  provies a relation
between the synthesized mass of $^{56}$Ni and the late-time bolometric
light  curve. Within  this  treatment, the  late-time  light curve  is
computed numerically using  a Monte Carlo algorithm,  which solves the
transport of photons, and the injection of energy by the $\gamma$ rays
produced  by the  decays  of  $^{56}$Ni and  $^{56}$Co.   In order  to
compare  with the  observed  duration of  fast  optical transients  we
compute  the   characteristic  timescales  of  the   light  curves  of
\cite{Gabriela2} as their width at  half the peak luminosity. The time
origin  is chosen  as the  point  at which  the dynamical  interaction
reaches the  peak temperature,  which occurs  shortly before  the peak
luminosity  is reached.  This  is close  enough  to the  observational
definition of the decay time ---  the time to decay one magnitude from
peak, or a factor of 2.5  in flux \citep{Kasliwal12a} --- to allow for
an approximate comparison between models and observations.

\begin{figure*}
\begin{center}
  \includegraphics[width=0.35\textwidth, trim=1.6cm 4.5cm 0.1cm 4.5cm, clip=true]{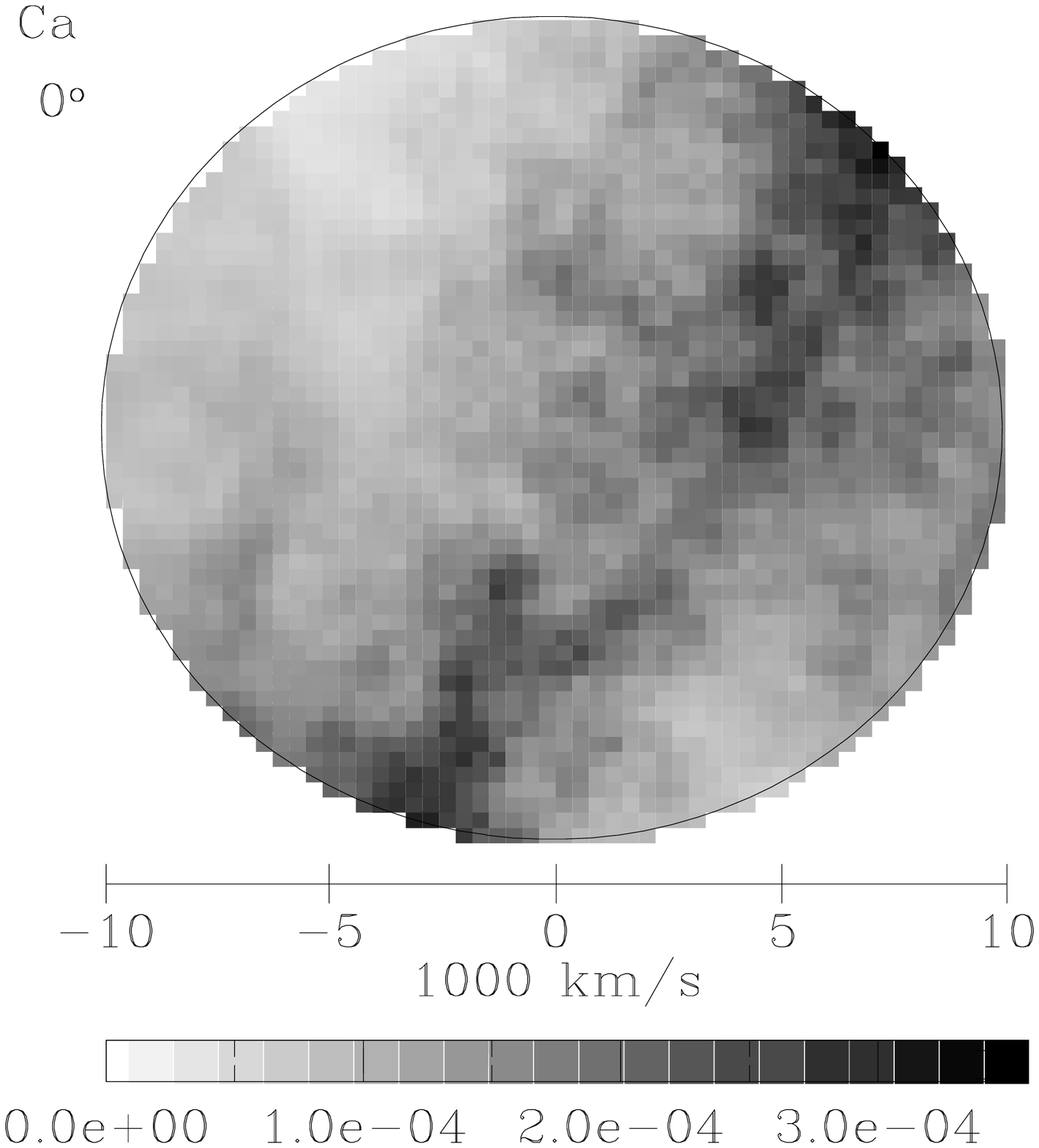}
  \includegraphics[width=0.35\textwidth, trim=1.6cm 4.5cm 0.1cm 4.5cm, clip=true]{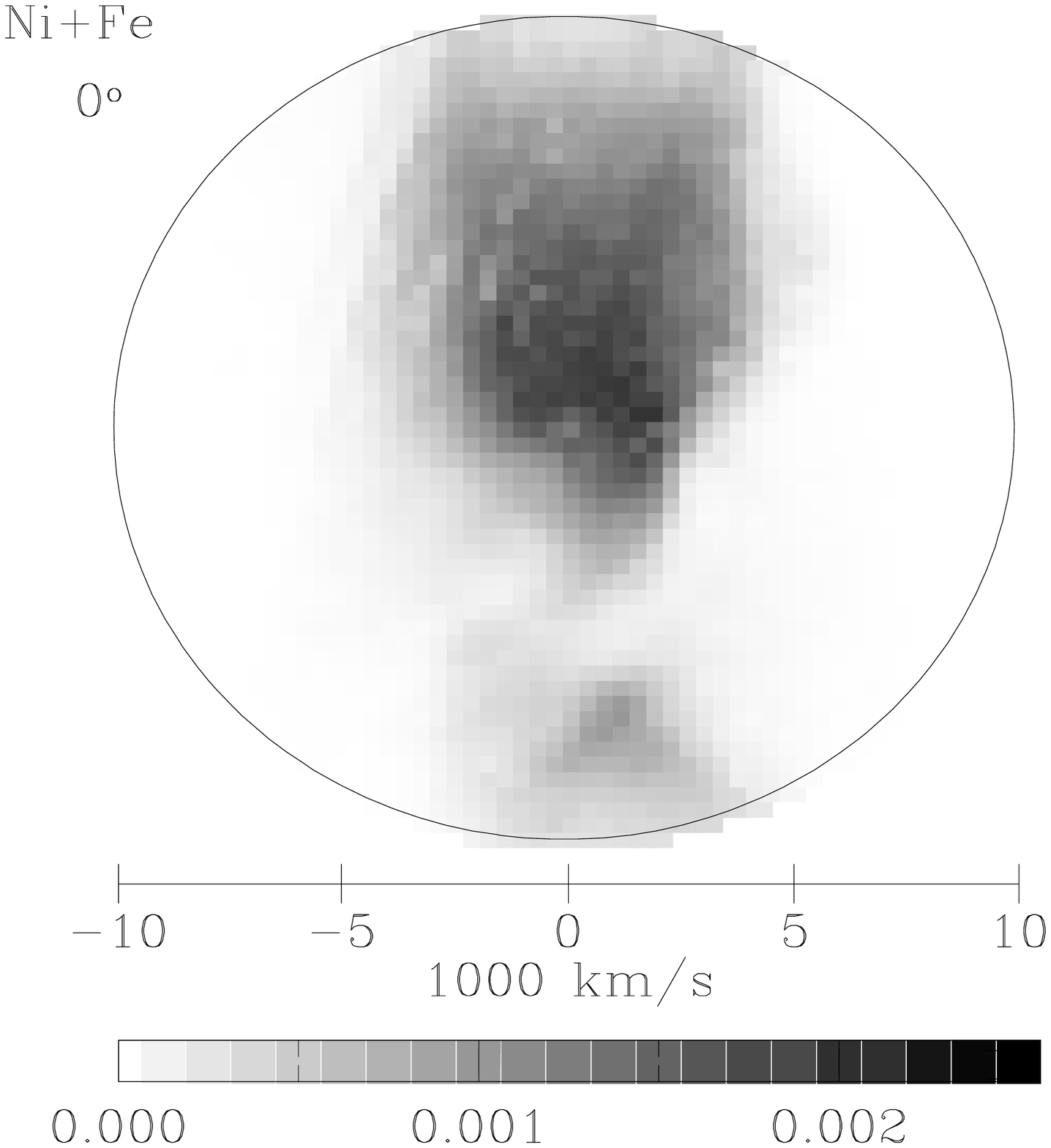}
  \includegraphics[width=0.35\textwidth, trim=1.6cm 4.5cm 0.1cm 4.5cm, clip=true]{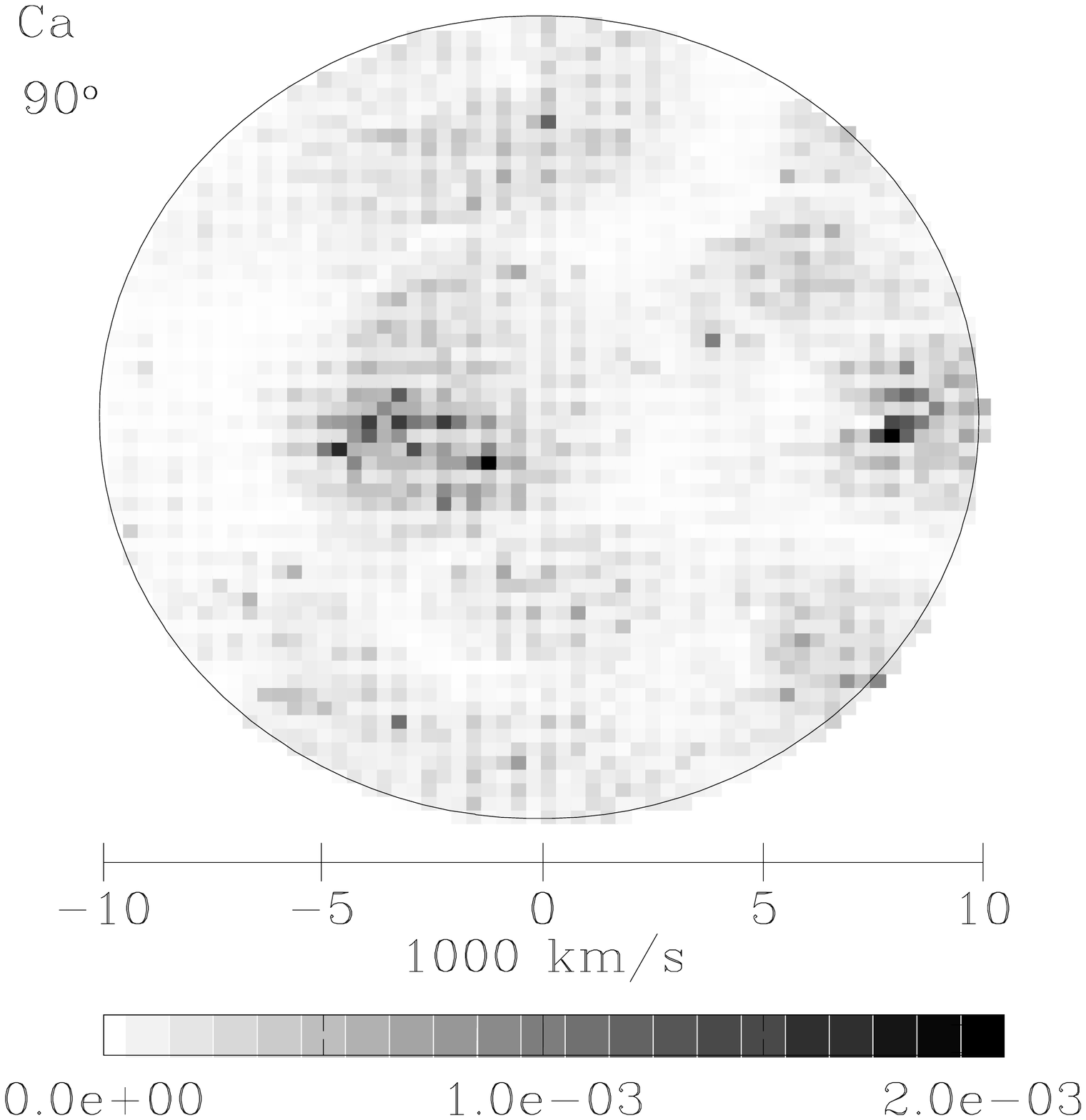}
  \includegraphics[width=0.35\textwidth, trim=1.6cm 4.5cm 0.1cm 4.5cm, clip=true]{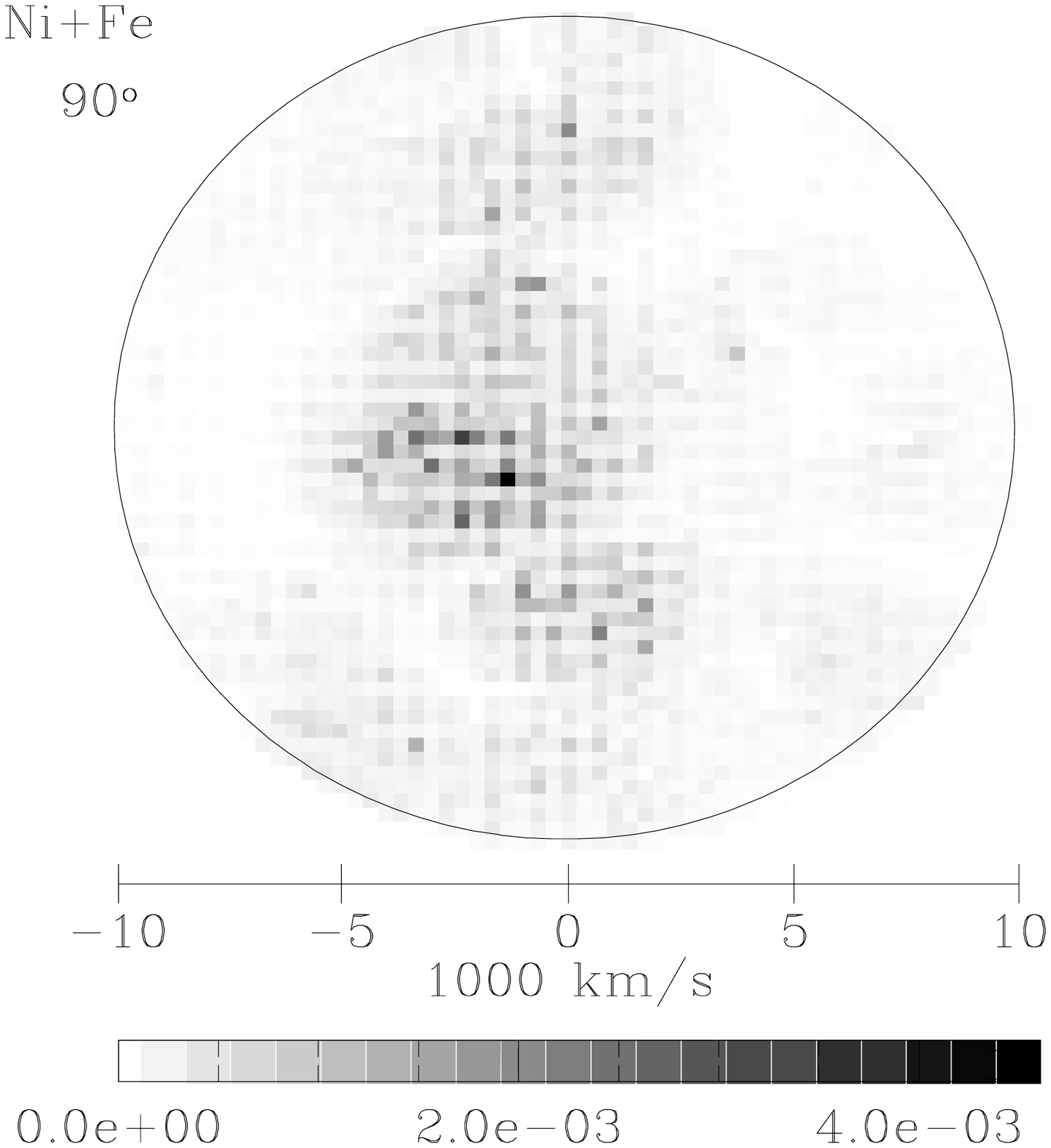}\\
\caption{Spatial distribution of the Ca (left panels) and Ni+Fe (right
  panels)  mass abundances  for  the  $0.4+0.4\,M_{\sun}$ merger  with
  $v_{\rm ini}=100$~km~s$^{-1}$ and  $\Delta y=0.4\,R_{\sun}$, and two
  inclinations, $0^\circ$ (top panels) and $90^\circ$ (bottom panels),
  as a function of the average expansion velocity.}
\label{fig:Ca}
\end{center}
\end{figure*}

In Fig.~\ref{fig:gap} the results  of these numerical calculations are
compared to the observational properties  of the several subclasses of
fast optical transients.  In particular,  the three main subclasses of
fast   transients  are   highlighted  in   magenta  (.Ia   SNe),  blue
(fast/bright transients)  and orange (Ca-rich transients).   The white
dwarfs  in  our  model  grid   are  composed  of  helium  (0.4~\msun),
carbon-oxygen (0.8 and 1.0~\msun) and oxygen-neon (1.2~\msun), and the
differences between models with identical components are driven mainly
by   the   initial   velocities   and  impact   parameters   ---   see
\cite{Gabriela1} for details.  As can be  seen in this figure, most of
our  model light  curves have  timescales of  $\sim 10$  days, clearly
faster than those  of normal \snia\ and slower than  those of .Ia SNe,
but comparable to both fast/bright  and Ca-rich transients.  When both
white  dwarfs are  relatively  massive, the  transients  can be  quite
bright, with peak luminosities  of $\sim 10^{44}$~erg~s$^{-1}$ for two
carbon-oxygen   cores,   and   $\sim  10^{43}$~erg~s$^{-1}$   for   an
interaction  involving  carbon-oxygen  and oxygen-neon  white  dwarfs.
These values  are comparable  with those  observed in  the fast/bright
transients  like   SN~1885  and  SN~1939B.   In   contrast,  dynamical
interactions  that involve  at least  one helium  white dwarf  produce
fainter transients with $10^{40}\la  L \la 10^{42}$~erg~s$^{-1}$.  The
upper end  of this range  is comparable  with the values  measured for
prototypical Ca-rich  transients like SN~2005E.  Finally,  some models
have lower peak luminosities, comparable  to those of SN~2008ha, while
other models have peak luminosities  one order of magnitude lower than
this.   At  present it  is  unclear  whether these  interactions  have
observational counterparts.

A more  detailed comparison between  our theoretical light  curves and
the observations  of two  fast and bright  transients ---  SN~1885 and
SN~1939B \citep{Perets2011} --- and  a Ca-rich transient --- PTF~10iuv
\citep{Kasliwal12b}   ---  is   shown  in   Fig.~\ref{fig:LC}.   These
comparisons are  necessarily qualitative  since, as  discussed before,
our  theoretical light  curves are  only approximate  for times  below
50~days. However, here we highlight the  behavior at late times, up to
250~days, where our calculations are  more reliable. In principle, the
luminosities in  our light curves  should not be directly  compared to
broadband  photometric observations  without some  sort of  bolometric
correction. This is less of a  concern for SN~1885 and SN~1939B, since
transients powered  by $^{56}$Ni decay,  like all normal  and peculiar
SN~Ia, and our white dwarf interaction models, tend to peak in the $B$
and $V$ band \citep{Ashall16},  where the historical observations were
taken.  There  might be a  larger correction for PTF~10iuv,  since the
light curve from \cite{Kasliwal12b} is in the $r$ band, but we include
the  comparison for  illustrative purposes  because this  is the  most
extended light curve available for  a Ca-rich transient.  Within these
limitations,  the  comparisons  in  Fig.~\ref{fig:LC}  show  that  the
lightcurves for  models involving carbon-oxygen and  oxygen-neon white
dwarfs do span the range  of peak luminosities and late-time evolution
for fast  and bright  transients. The match  between SN~1939B  and the
carbon-oxygen and  oxygen-neon models is particularly  good, while the
late time decay of SN~1885 seems faster than the models, though not by
much. The  late time evolution  of the Ca-rich transient  PTF~10iuv is
similar  to that  of  the models  which most  closely  match its  peak
luminosity, which include the interactions of two white dwarfs made of
helium and an oxygen-neon white dwarf with a white dwarf with a helium
core.

\section{Calcium yields and spatial distribution}
\label{sec:X}

Table~\ref{tab:calcium} lists the  Ca, Ni, and Fe yields  for the runs
of \cite{Gabriela1} that have an  explosive outcome --- see Table~6 of
\cite{Gabriela1}.  In  this table  we also  show the  initial relative
velocity  ($v_{\rm   ini}$)  and   separation  ($\Delta  y$)   of  the
interacting white dwarfs. Our dynamical white dwarf interaction models
synthesize between  $\sim0.001$ and $\sim0.06$~\msun\ of  Ca, with the
highest yields  corresponding to  the 0.8+1.0~\msun\  simulation.  For
the  interactions involving  at least  one helium  white dwarf,  which
provide the best  match for the properties of  Ca-rich transients, the
Ca yield  can be as high  as $\sim0.02$~\msun.  This is  comparable to
the  0.03~\msun\  of   Ca  in  the  model   by  \cite{Waldman11}  that
\cite{Dessart15} matched to the properties of Ca-rich transients.

The spatial distribution of Ca and Fe+Ni is shown in Fig.~\ref{fig:Ca}
for a representative model.  Namely, a $0.4+0.4\, M_{\sun}$ collision,
with initial separation $\Delta  y=0.3\,R_{\sun}$ and initial velocity
$v_{\rm  ini}=75$~km~s$^{-1}$.    These  two-dimensional  distribution
plots, shown  for two  different planes  in the  model ejecta,  can be
directly compared  to the \textit{HST} narrow-band  absorption maps of
SN~1885  ---   Fig.~8  in  \cite{Fesen2015}.   There   are  remarkable
qualitative  similarities between  these observed  maps and  the model
results.   The  distribution of  the  elements  in velocity  space  is
similarly  compact,  and  shows  a large  degree  of  assimmetry.   In
particular,   little   material   has    velocities   in   excess   of
10,000~km~s$^{-1}$.  Moreover, our sample model also shows very clumpy
and asymmetric ejecta, just like the SN~1885 images.  Specifically, in
the Ca map  the $0^\circ$ projection shows a  prominent band extending
from  $\sim -5,000$  to  $\sim 7,000$~km~s$^{-1}$,  whereas the  Fe+Ni
distribution shows two  prominent clumps at low  velocities, being the
top one more evident. All these features have roughly the same angular
size as  the ones  seen in  the SN~1885  map.  In  the Fe+Ni  map, the
$90^\circ$  projection  shows  some   filamentary  structure  that  is
reminiscent  of, albeit  less marked  than, the  one seen  in SN~1885.
However, we emphasize that the detailed structure of the debris region
depends sensitively  on the  orientation with respect  to the  line of
sight, and  that for  some inclinations  the filamentary  structure is
more  evident.   Almost  the same  can  be  said  for  the Ca  map  at
$90^\circ$, for  which two  clear structures can  be seen  at opposite
directions, having velocities of  $\sim -4,000$~km~s$^{-1}$, and $\sim
7,000$~km~s$^{-1}$, respectively.   Note that  \cite{Fesen2015} argued
that SN~1885 was a sub-luminous Type~Ia supernova with an ejected mass
close to the Chandrasekhar limit.  However, \cite{Perets2011} based on
the lack  of X-ray  emission 130~yr  after the  explosion, and  on the
shape of the light curve, did not find clear evidence for this. A more
detailed comparison between model  lightcurves and the historical data
for SN~1885,  as well as a  statistical analysis of the  HST images in
the  context of  multi-dimensional explosion  models will  settle this
issue.  We defer these comparisons to future work.

The   total  mass   of   Ca  for   SN~2005E   is  $0.135\,   M_{\sun}$
\citep{Perets10}, and  substantially smaller, $0.054\,  M_{\sun}$, for
PTF~10iuv  \citep{Kasliwal12b}.    The  Ca  masses  obtained   in  our
simulations --- see Table~\ref{tab:calcium}  --- are somewhat smaller,
albeit comparable to  those of Ca-rich transients.   For instance, for
the simulation presented in Fig.~\ref{fig:Ca}, which shows the spatial
distribution of Ca  for the collision of two helium  white dwarfs, the
total Ca  mass is $\sim  0.02\, M_{\sun}$. Nevertheless,  a cautionary
note  is in  order here,  since the  observed Ca  masses are  somewhat
uncertain, as  they are  sensitive to the  adopted temperature  of the
nebular spectrum.  In fact, a change of the nebular temperature can
result in a sizable change in the derived masses \citep{Kasliwal12b}.

\section{Summary and conclusions}
\label{sec:concl}

We have  explored the fundamental properties  (light curve timescales,
peak  luminosities, and  nucleosynthetic yields)  of a  grid of  white
dwarf collision models by \cite{Gabriela1},  and we have compared them
to the known  properties of fast transients. These  collisions are not
frequent,   except   in   dense  stellar   environments,   and   their
observational counterparts  have not been clearly  identified yet.  We
find that collisions in which at least one of the interacting stars is
a helium white  dwarf match the properties of  Ca-rich transients like
SN~2005E and unclassified transients  like SN~2008ha, while collisions
that have more massive  components (carbon-oxygen or oxygen-neon white
dwarfs)  match  those  of  fast/bright  transients,  like  SN~1885  or
SN~1939B.  In  particular, we  find that the  timescales of  the light
curves of those  collisions in which one of the  members of the system
is  a white  dwarf  with a  helium core  are  compatible with  Ca-rich
transients. Also,  the mass of Ca  in the debris region  formed around
the more massive  white dwarf, which remains almost  intact during the
dynamical  interaction,  is  comparable  to  that  measured  in  these
transients. The  velocities of  the material in  this region  are also
consistent  with the  measured  velocities of  the  ejecta of  Ca-rich
transients.   Moreover,  the  spatial  distribution of  Ca-,  Ni-  and
Fe-rich material  resembles the observed distributions  around Ca-rich
transients.  Finally,  it is worth  mentioning that all of  our models
produce light curves that are faster  than those of \sneia, but slower
than those of  .Ia SNe.  Nevertheless, we emphasize  that more studies
are  needed to  confirm or  rule  out whether  these interactions  are
indeed  the  progenitors  of  at  least a  fraction  of  fast  optical
transients.   Looking to  the  future, further  work  should focus  on
detailed radiative  transfer modeling of  light curves and  spectra to
compare  with extant  and  future observations.   This  would allow  a
better comparison with observations.

\section*{Acknowledgements}

This work was partially funded by the MINECO grant AYA2014-59084-P and
by the  AGAUR (EG-B).  CB acknowledges support  from grants  NASA ADAP
NNX15AM03G S01 and NSF/AST-1412980. We acknowledge the useful comments
of our referee, which helped in  improving the original version of the
paper.

\bibliographystyle{mnras}
\bibliography{CaTs}

\bsp	
\label{lastpage}

\end{document}